\documentstyle[sprocl,english]{article}

\def\beq{\begin{equation}}
\def\eeq{\end{equation}}
\def\6{\langle}
\def\9{\rangle}
\def\half{\mbox{$1\over2$}}
\def\bn{\mbox{\boldmath$n$}}
\def\nJ{\mbox{\boldmath$n\cdot J$}}
\def\dtp{d_{\theta\phi}}
\def\DD{{\cal D{}}}
\def\ptp{\psi\theta\phi}
\def\xyz{\xi\eta\zeta}
\def\mnr{\mu\nu\rho}
\def\abc{\alpha\beta\gamma}
\def\0{\otimes}
\def\1{\mbox{1\hskip-.25em l}}

\begin{document}

\title{Unspeakable quantum information}

\author{Asher Peres and Petra F. Scudo}
\address{Department of Physics, Technion---Israel Institute of
Technology, Haifa, Israel}

\maketitle
\begin{abstract}
No verbal explanation can indicate a direction in space or the
orientation of a coordinate system. Only material objects can do it.
In this article we consider the use of a set of spin-\half\ particles
in an entangled state for indicating a direction, or a hydrogen atom in
a Rydberg state for transmitting a Cartesian frame. Optimal strategies
are derived for the emission and detection of the quantum signals.

\end{abstract}

\bigskip
\section{Formulation of the problem}
The role of a dictionary is to define unknown words by means of known
ones. However there are terms, like left or right, which cannot be
explained in this way. In the absence of a formal definition, material
objects must be used to illustrate these terms: for example, we may
say that the human liver is on the right side. Likewise, the sign
of helicity may be referred to the DNA structure, or to the properties
of weak interactions.

Here we consider some cases where information cannot be explained
verbally. The simplest one  is when the emitter (conventionally called
Alice) wants to indicate to the receiver (Bob) a direction in space.
If they have a common coordinate system to which they can refer, or
if they can create one by observing distant fixed stars, Alice simply
communicates to Bob the components of a unit vector \bn\ along
that direction, or its spherical coordinates $\theta$ and $\phi$. But
if no common coordinate system has been established, all she can do is
to send a real physical object, such as a gyroscope, whose orientation
is deemed stable.

\section{Quantum transmission of a spatial direction}
In the quantum world, the role of the gyroscope is played by a system
with large spin. For example, Alice can send angular momentum eigenstates
satisfying $\nJ|\psi\9=j|\psi\9$. This is essentially the solution
proposed by Massar and Popescu \cite{mp} who took $N$ parallel spins,
polarized along~\bn. The fidelity of the transmission is usually defined
as

\beq F=\6\cos^2(\chi/2)\9=(1+\6\cos\chi\9)/2, \eeq
where $\chi$ is the angle between the true \bn\ and the direction
indicated by Bob's measurement. The physical meaning of $F$ is that
$1-F=\6\sin^2(\chi/2)\9$ is the mean square error of the measurement,
if the error is defined as $\sin(\chi/2)$.~\cite{helstrom} The
experimenter's aim, minimizing the mean square error, is the same as
maximizing fidelity. We can of course define ``error'' in a different
way, and then fidelity becomes a different function of~$\chi$ and
optimization leads to different results \cite{m}. Here, we shall take
Eq.~(1) as the definition of fidelity.

Massar and Popescu showed that for parallel spins, \mbox{$1-F=1/(N+2)$.}
It then came as a surprise that for $N=2$, parallel spins were not
the optimal signal, and a slightly higher fidelity resulted from the
use of opposite spins~\cite{gp}. The intuitive reason given for this
result was the use of a larger Hilbert space (four dimensions instead
of three). This raises the question what is the most efficient signal
state for $N$ spins, whose Hilbert space has $2^N$ dimensions. Will $F$
approach~1 exponentially?  Actually, the optimal result is a quadratic
approach, as shown below.

Our first task is to devise Bob's measuring method, whose mathematical
representation is a positive operator-valued measure (POVM)
\cite{qt,maxwell}.  For any unit vector \bn, not necessarily Alice's
direction, let $|j,m(\bn)\9\equiv|j,m(\theta,\phi)\9$ denote the
coherent angular momentum state \cite{perel} that satisfies

\beq \mbox{\boldmath$J$}^2\,|j,m(\bn)\9=j(j+1)\,|j,m(\bn)\9, \eeq
and
\beq \nJ\,|j,m(\bn)\9= m\,|j,m(\bn)\9. \eeq
We then have \cite{perel}

\beq (2j+1)\int\dtp\,|j,m(\theta,\phi)\9\6j,m(\theta,\phi)|
  =\1_j, \label{1j}\eeq
where 

\beq \dtp:=\sin\theta d\theta\,d\phi/4\pi, \eeq
and $\1_j$ is the projection operator over the $(2j+1)$-dimensional
subspace spanned by the vectors $|j,m(\theta,\phi)\9$. If $N=2$, so that
$j$ is 0 or 1, the two resulting subspaces span the whole 4-dimensional
Hilbert space. For higher $N$, all the rotation group representations
with $j<N/2$ occur more than once.  We then have, if we take each $j$
only once, from 0 or \half\ to $N/2$,

\beq \sum(2j+1)={(N+2)^2\over4}\quad\mbox{or}
  \quad {(N+1)(N+3)\over4},\eeq
for even or odd $N$, respectively. For large $N$, the dimensionality of
the accessible Hilbert space tends to $N^2/4$, and this appears to be
the reason that the optimal result for $1-F$ is quadratic in $N$, not
exponential.

We now turn to the construction of Bob's POVM \cite{qt}.  Let $\rho$
denote the initial state of the physical system that is measured. All
these input states span a subspace of Hilbert space.  Let \1 denote
the projection operator on that subspace. A POVM is a set of positive
operators $E_\mu$ which sum up to \1. The index $\mu$ is just a
label for the outcome of the measuring process. The probability of
outcome $\mu$ is $\mbox{tr}(\rho E_\mu)$.  In the present case, $\mu$
stands for the pair of angles $\theta\phi$ that are indicated by Bob's
measurement. If we want a high accuracy, these output angles should have
many different values, spread over the unit sphere \cite{polyhed}. For
example, the components of a continuous POVM, as in Eq.~(\ref{1j}), are
given by

\beq E_{\theta\phi}=(2j+1)\,\dtp\,|j,m(\theta,\phi)\9\6j,m(\theta,\phi)|.
 \eeq
Such a POVM with $m=j$ corresponds to the method of Massar and
Popescu~\cite{mp}. The choice $m=j$ is not optimal. As shown by Gisin
and Popescu \cite{gp} for the case $N=2$, signal states with opposite
spins give a higher fidelity. With our present notations, these states
are $(|0,0\9+|1,0(\bn)\9)/\sqrt2$. They involve two values of $j$,
but a single value of $m$, namely~0.

One possibility to include several values of $j$ in a POVM is to take a
sum of expressions like (\ref{1j}). This brings no advantage, because a
convex combination of POVMs cannot yield more information than the best
one of them \cite{cnvx}. Optimal POVM components are usually assumed
to have rank one. In the present case, each one of them should include
all relevant $j$:

\beq E_{\theta\phi}:=\dtp\,|\theta,\phi\9\6\theta,\phi|, \eeq
where
\beq |\theta,\phi\9:=\sum_{j=m}^{N/2}\sqrt{2j+1}\;|j,m(\theta,\phi)\9.
 \label{thphi}\eeq

To verify that this is indeed a POVM, we note that in 
$\int\!E_{\theta\phi}$ there are diagonal terms $(2j+1)|j,m(\theta,\phi)\9
\6j,m(\theta,\phi)|$, which give $\1_j$, owing to Eq.~(\ref{1j}).
The off-diagonal terms with $j_1\neq j_2$ vanish, as can be seen by
taking their matrix elements between $\6j_1,m_1|$ and $|j_2,m_2\9$
in the standard basis where $J_z$ is diagonal. We have \cite{edm}

\beq \6j_2,m(\theta,\phi)|j_2,m_2\9=
  {\cal D}^{(j_2)}_{mm_2}(\psi\theta\phi), \eeq
with a similar (complex conjugate) expression for
$\6j_1,m_1|j_1,m(\theta,\phi)\9$. The rotation matrices $\cal D$ are
explicitly given by

\beq {\cal D}^{(j_2)}_{mm_2}(\psi\theta\phi)=
  e^{im\psi}\,d^{(j_2)}_{mm_2}(\theta)\,e^{im_2\phi}, \eeq
where the Euler angle $\psi$ is related to an arbitrary phase which is
implicit in the definition of $|j,m(\theta,\phi)\9$. Note that
a single value of $m$ occurs in all the components of the vectors
$|\theta,\phi\9$ in Eq.~(\ref{thphi}), so that the undefined phases
$e^{\pm im\psi}$ mutually cancel. It then follows from Eq.~(4.6.1) of
Edmonds \cite{edm} that all the off-diagonal matrix elements of
$\int\!E_{\theta\phi}$ vanish, so that we indeed have a POVM~\cite{mix}.

While Bob's optimal POVM is essentially unique in the Hilbert space that
we have chosen, Alice's signal state, which is

\beq |A\9=\sum_{j=m}^{N/2} c_j\,|j,m(\bn)\9, \label{alice} \eeq
contains unknown coefficients $c_j$. The latter are normalized,

\beq \sum_{j=m}^{N/2} |c_j|^2 = 1, \label{norm} \eeq
but still have to be optimized. 

The probability of detection of the pair of angles $\theta\phi$,
indicated by the POVM component $E_{\theta\phi}$, is

\beq \6A|E_{\theta\phi}|A\9=\dtp\left|\sum_{j=m}^{N/2}c_j\sqrt{2j+1}\,
  \6j,m(\theta,\phi)|j,m(\bn)\9\right|^2. \label{AEA} \eeq
We have \cite{perel}

\beq \6j,m(\theta,\phi)|j,m(\bn)\9=e^{i\eta}\,d^{(j)}_{mm}(\chi), \eeq
where $\chi$ is the angle between the directions \bn\ and
$\theta\phi$, and the phase $e^{i\eta}$ is related to the arbitrary
phases which are implicit in the definitions of the state vectors in
(\theequation).  The important point is that $e^{i\eta}$ does not depend
on $j$ and therefore is eliminated when we take the absolute value of
the sum in Eq.~(\ref{AEA}). Explicitly, we have

\beq d^{(j)}_{mm}(\chi)=\cos^{2m}(\chi/2)\,P^{(0,2m)}_{j-m}(\cos\chi),\eeq
where $P^{(a,b)}_n(x)$ is a Jacobi polynomial \cite{perel,edm}. We
shall write $x=\cos\chi$ for brevity, so that the fidelity is

\beq F=(1+\6x\9)/2. \eeq
Our problem is to find the coefficients $c_j$ that maximize $\6x\9$.
Owing to rotational symmetry, we can assume that Alice's
direction \bn\ points toward the $z$-axis, so that $\dtp$ can be
replaced by $dx/2$ after having performed the integration over $\phi$. We
thus obtain

\beq \6x\9=\half\int^1_{-1}xdx\left|\sum_{j=m}^{N/2}c_j\sqrt{2j+1}
 \left({1+x\over2}\right)^mP^{(0,2m)}_{j-m}(x)\right|^2. \eeq

This integral can be evaluated explicitly by using the orthogonality and
recurrence relations for Jacobi polynomials \cite{rain}. The result is

\beq \6x\9=\sum_{j,k} c^*_j\,c_k\,A_{jk}, \eeq
where $A_{jk}$ is a real symmetric matrix, whose only nonvanishing
elements are

\beq A_{jj}=m^2/[j(j+1)], \eeq
and
\beq A_{j,j-1}=A_{j-1,j}=(j^2-m^2)/j\sqrt{4j^2-1}. \eeq
The optimal coefficients $c_j$ are the components of the eigenvector
of $A_{jk}$ that corresponds to the largest eigenvalue, and the latter
is $\6x\9$ itself. For $m=0$ (which is the optimal choice) and large $N$, 
we find that

\beq 1-F\to[2.4048/(N+3)]^2, \label{infi1}\eeq
where the numerator is the first zero of the Bessel function
$J_0$.\cite{bbbmt}  The right hand side ought to be compared to the
result of Massar and Popescu \cite{mp}, which was $1/(N+2)$. A detailed
comparison of both methods is illustrated by the figure which appears in
a previous article \cite{petra}.

\section{Transmission of a Cartesiam frame}
Next, we consider the quantum transmission of a complete
Cartesian frame. If many spins are available, a simple possibility
would be for Alice to use half of them for indicating her $x$ axis and
the other half for her $y$ axis. However, the two directions found
by Bob may not then be exactly perpendicular, because separate
transmissions have independent errors due to limited angular
resolution. Some adjustment will be needed to obtain Bob's best
estimates for the $x$ and $y$ axes, before he can infer from them his
guess of Alice's $z$ direction. This method is not optimal, and it is
obviously not possible to proceed in this way if a {\it single\/} quantum
messenger is available. Here we shall show how a single hydrogen atom
(formally, a spinless particle in a Coulomb potential) can transmit a
complete frame.

Consider the $n$-th energy level of that atom (a Rydberg state). Its
degeneracy is $d=n^2$ because the total angular momentum may take
values $j=0,\cdots,n-1$, and for each one of them $m=-j,\cdots,j$. Alice
indicates her $xyz$ axes by sending the atom in a state

\beq |A\9=\sum_{j=0}^{n-1}\;\sum_{m=-j}^j a_{jm}\,|j,m\9,\label{A}\eeq
with normalized coefficients $a_{jm}$ that will be specified below. Bob
then performs a covariant measurement \cite{holevo} in order to evaluate
the Euler angles $\ptp$ that would rotate his own $xyz$ axes into a
position parallel to Alice's axes. Bob's detectors (ideally, there is
an infinite number of them \cite{finite}) have labels $\ptp$ and the
mathematical representation of his apparatus is as usual a POVM. In
the present case, we have a resolution of identity by a set of positive
operators:

\beq  \int d_{\ptp}\,E(\ptp)=\1, \label{povm}\eeq
where $d_{\ptp}\equiv\sin\theta d\psi d\theta d\phi/8\pi^2$ is the
$SO(3)$ Haar measure for Euler angles \cite{edm}, and
$E(\ptp)=|\ptp\9\6\ptp|$. The vectors $|\ptp\9$ will be specified
below. The probability that the detector labelled $\ptp$ is excited is
given by

\beq P(\ptp)= \6A|d_{\ptp} E(\ptp)|A\9=
  d_{\ptp}\,|\6A|\ptp\9|^2. \label{prob} \eeq
Our task is to construct vectors $|\ptp\9$ such that Eq.~(\ref{povm})
is satisfied (that is, the probabilities sum up to one) and Bob's
expected error is minimal.

Generalizing the method of the preceding section, we define a fiducial
vector for Bob,

\beq |B\9=\sum_{j=0}^{n-1}\sqrt{2j+1}\sum_{m=-j}^j b_{jm}\,|j,m\9,
  \label{B} \eeq
where the coefficients $b_{jm}$ are normalized for each $j$ separately:

\beq \sum_{m=-j}^j\,|b_{jm}|^2=1\qquad\forall j. \label{b} \eeq
For the preceding problem, a single value of $m$ was used; here we
need all the values. Note that Eq.~(\ref{A}) was written with Alice's
notations ($m$ is the angular momentum along her $z$ axis), while
Eq.~(\ref{B}) is written with Bob's notations ($m$ refers to his $z$
axis). This issue will be dealt with later.

We now define

\beq |\ptp\9=U(\ptp)\,|B\9, \eeq
where $U(\ptp)$ is the unitary operator for a rotation by Euler angles
$\ptp$. Note that since $|B\9$ is a direct sum of vectors, one for each
value of $j$, then likewise $U(\ptp)$ is a direct sum with one term
for each irreducible representation,

\beq U(\ptp)=\sum_j\oplus\DD^{(j)}(\ptp), \eeq
where the $\DD^{(j)}(\ptp)$ are the usual irreducible unitary rotation
matrices \cite{edm}.  To prove that Eq.~(\ref{povm}) is satified, we
note that its left hand side is invariant if multiplied by $U(\mnr)$
on the left and $U(\mnr)^\dagger$ on the right, for any arbitrary Euler
angles $\mnr$ (because these unitary matrices represent group elements
and therefore have the group multiplication properties) \cite{perel}. It
then follows from a generalization of Schur's lemma \cite{wigner} that
the left hand side of (\ref{povm}) is a direct sum of {\it unit\/}
matrices, owing to the presence of the factor $(2j+1)$ which is the
dimensionality of the corresponding irreducible representation. Therefore
Eq.~(\ref{povm}) is satisfied.

The detection probability in Eq.~(\ref{prob}) can thus be written as
$P(\ptp)= d_{\ptp}\,|\6A|U(\ptp)|B\9|^2.$ To compute this expression
explicitly, we must use consistent notations for $|A\9$ and
$|B\9$ --- recall that Eq.~(\ref{A}) was written in Alice's basis, and
Eq.~(\ref{B}) in Bob's basis. It is easier to rewrite Alice's vector
$|A\9$ in Bob's language. For this we have to introduce the Euler angles
$\xyz$ that rotate Bob's $xyz$ axes into Alice's axes (that is, $\xyz$
are the true, but unknown values of the angles $\ptp$ sought by Bob).
The unitary matrix $U(\xyz)$ represents an {\it active\/} transformation
of Bob's vectors into Alice's. Therefore, $U(\xyz)^\dagger$ is the {\it
passive\/} transformation \cite{qt} from Bob's notations to those of
Alice, and $U(\xyz)$ is the corresponding transformation from Alice's
notations to Bob's. Written in Bob's notations, Alice's vector $|A\9$
becomes $U(\xyz)|A\9$ so that, in Eq.~(\ref{prob}), $\6A|$ becomes
$\6A|U(\xyz)^\dagger$. Let us therefore define

\beq U(\abc)=U(\xyz)^\dagger\,U(\ptp). \eeq
The Euler angles $\abc$ have the effect of rotating Bob's Cartesian
frame into his {\it estimate\/} of Alice's frame, and then rotating
back the result by the {\it true\/} rotation from Alice's to Bob's
frame. That is, the angles $\abc$ indicate Bob's measurement error.
The probability of that error is

\beq P(\abc)= d_{\abc}\,|\6A|U(\abc)|B\9|^2, \label{errprob}\eeq
where $d_{\abc}=\sin\beta d\alpha d\beta d\gamma/8\pi^2$. Note that
in the above equation $|A\9$ is written with Alice's notations as in
(\ref{A}), and $|B\9$ with Bob's notations as in (\ref{B}). 

Of course Bob cannot know the values of $\abc$. His measurement only
yields some value for $\ptp$. The following calculation that employs
$\abc$ has the sole purpose of estimating the expected accuracy of the
transmission (which does not depend on the result $\ptp$ owing to
rotational symmetry).

We must now choose a suitable quantitative criterion for that accuracy.
When a single direction is considered, it is convenient to define the
error \cite{helstrom} as $\sin(\omega/2)$, where $\omega$ is the angle
between the true direction and the one estimated by Bob. The mean square
error is

\beq \6\sin^2(\omega/2)\9=(1-\6\cos\omega\9)/2=1-F, \eeq
where $F$ is the fidelity, as defined above. When
we consider a Cartesian frame, we likewise define fidelities for
each axis. Note that $\cos\omega_k$ (for the $k$-th axis) is given
by the corresponding diagonal element of the orthogonal (classical)
rotation matrix.  Explicitly, we have \cite{goldstein}

\beq \cos\omega_z=\cos\beta, \label{z} \eeq and
\beq \cos\omega_x+\cos\omega_y=(1+\cos\beta)\,\cos(\alpha+\gamma), 
 \label{xy} \eeq
whence, by Euler's theorem,

\beq \cos\omega_x+\cos\omega_y+\cos\omega_z=1+2\cos{\mit\Omega},
 \label{xyz} \eeq
where $\mit\Omega$ has a simple physical meaning: it is the angle for
carrying one frame into the other by a single rotation.

The expectation values of the above expressions are obtained with the
help of Eq.~(\ref{errprob}):

\beq \6f(\abc)\9=\int d_{\abc}\,|\6A|U(\abc)|B\9|^2\,f(\abc), \eeq
where, explicitly,

\beq \6A|U(\abc)|B\9=\sum_{j,m,r}\;a_{jm}^*\,b_{jr}
  \6j,m|\DD^{(j)}(\abc)|j,r\9. \eeq
The unitary irreducible rotation matrices $\DD^{(j)}(\abc)$ have
components \cite{edm}
\beq \6j,m|\DD^{(j)}(\abc)|j,r\9=
  e^{i(m\alpha+r\gamma)}\,d^{(j)}_{mr}(\beta), \eeq
where the $d^{(j)}_{mr}(\beta)$ can be expressed in terms of Jacobi
polynomials. Collecting all these terms, we finally obtain, after many
tedious analytical integrations over products of Jacobi
polynomials \cite{rain},

\beq \6f(\abc)\9=\sum f_{jkmnrs}\,a_{jm}^*\,b_{jr}\,a_{kn}\,b_{ks}^*,
 \label{fabab} \eeq
where the numerical coefficients $f_{jkmnrs}$ depend on our choice of
$f(\abc)$ in Eqs.~(\ref{z}--\ref{xyz}). The problem is to optimize the
components $a_{jm}$ (normalized to 1), and $b_{jm}$ satisfying the
constraints (\ref{b}), so as to maximize the above expression. For
further use, it is convenient to define a matrix

\beq M_{jm,kn}=\sum_{r,s}f_{jkmnrs}\,b_{jr}\,b^*_{ks},\label{bilin}\eeq
so that 

\beq \6f(\abc)\9=\sum M_{jm,kn}\,a^*_{jm}\,a_{kn}=\6A|M|A\9.
 \label{Maa} \eeq

First consider a simple case: to transfer only the $z$ axis, we wish to
maximize $\6\cos\beta\9$. An explicit calculation yields

\beq f_{jkmnrs}=\delta_{mn}\,\delta_{rs}\,g_{jk}.\label{fddg} \eeq 
The matrix $g_{jk}$ which is defined by the above equation has
nonvanishing elements

\beq g_{jj}=ns/[j(j+1)], \eeq  and
\beq g_{j,j-1}=g_{j-1,j}={1\over j}
 \sqrt{{(j^2-n^2)(j^2-s^2)\over4j^2-1}}. \eeq
The $\delta_{mn}$ term in (\ref{fddg}) implies that, for any choice of
Bob's fiducial vector $|B\9$, the matrix $M$ in (\ref{Maa}) is
block-diagonal, with one block for each value of $m$. The optimization
of Alice's signal results from the highest eigenvalue of that matrix.
This is the highest eigenvalue of one of the blocks, so that a single
value of $m$ is actually needed. A similar (slightly more complicated)
argument applies if Alice's vector is given and we optimize Bob's
fiducial vector. This result proves the correctness of the intuitive
assumption that was made in our solution of the first problem, for which
a single value of $m$ was used. It was then found that when $m=0$ (the
optimal value) the expected error asymptotically behaves as $1.446/d$,
where $d$ is the effective number of Hilbert space dimensions. In the
present case, $d=(j_{{\rm max}}+1)^2$, which is the degeneracy of the
\mbox{$n$-th} energy level.

If we want to transfer two axes, we use Eq.~(\ref{xy}) and calculate
the matrix elements for $\6(1+\cos\beta)\cos{(\alpha+\gamma)}\9$. (It is
curious that they are simpler than those for $\6\cos{(\alpha+\gamma)}\9$
alone.) We obtain

\beq f_{jkmnrs}=\delta_{m,n-1}\,\delta_{r,s-1}\,h_{jk}+
 \delta_{n,m-1}\,\delta_{s,r-1}\,h_{kj}, \label{h} \eeq
where the nonvanishing elements of $h_{jk}$ are

\begin{eqnarray}
h_{jj}&=&\frac{[(j-n+1)(j+n)(j-s+1)(j+s)]^{1/2}}{2j(j+1)}, \\
h_{j,j-1}&=&\frac{[(j-n+1)(j-n)(j-s+1)(j-s)]^{1/2}}
 {2j(4j^2-1)^{1/2}}, \\
h_{j-1,j}&=&\frac{[(j+n-1)(j+n)(j+s-1)(j+s)]^{1/2}}
 {2j(4j^2-1)^{1/2}}. \end{eqnarray}
Note that the $h_{jk}$ matrix, whose elements depend on $n$ and $s$,
is not symmetric (while $g_{jk}$ was). This is because it comes from
the operator $e^{i(\alpha+\gamma)}$ which is not Hermitian. However
the two terms of (\ref{h}) together, which corresponds to
$\cos(\alpha+\gamma)$, have all the symmetries required by the other
terms $a_{jm}^*\,b_{jr}\,a_{kn}\,b_{ks}^*$ in Eq.~(\ref{fabab}).
Finally, if we wish to optimize directly the three Cartesian axes
(without losing accuracy by inferring $z$ from the approximate knowledge
of $x$ and $y$) we use all the terms of (\ref{xyz}), that is, both
those of (\ref{fddg}) and of (\ref{h}).

It now remains to find the vectors $|A\9$ and $|B\9$ that minimize
the transmission error. For small values of $j$, we used Powell's
method \cite{nr} without imposing any restrictions on $|A\9$ and $|B\9$
other than their normalization conditions. As intuitively expected,
we found that the optimal vectors satisfy

\beq b_{jm}=a_{jm}\left(\sum_n|a_{jn}|^2\right)^{-1/2},
  \qquad\forall j. \eeq
This means that Bob's vector should look as much as possible as Alice's
signal, subject to the restrictions imposed by the constraint~(\ref{b}).

Taking this property for granted is the key to a more efficient
optimization method, as follows: assume any $b_{jm}$, so that the
bilinear form (\ref{bilin}) is known. Find its highest eigenvalue and the
corresponding eigenvector $a_{jm}$. From the latter, get new components
$b_{jm}$ by means of (\theequation), and repeat the process until it
converges (actually, a few iterations are enough). Quantitative results 
are displayed in the figure of a preceding article \cite{euler}.
That figure shows that there is little advantage in optimizing only
two axes, if for any reason the third axis is deemed less important. If
the three axes are simultaneously optimized, it can be shown \cite{bbm}
that the mean square error tends asymptotically to $1/\sqrt{d}$.

It is not surprising that this result is weaker than the one for a
single axis, which was $1.446/d$. The obvious reason is that we are now
transmitting a three-dimensional rotation operation that can be applied
to any number of directions, not only to three orthogonal axes. Indeed,
consider any set of unit vectors ${\bf e}^\mu_m$, where $m=1,2,3$, and
$\mu$ is a label for identifying the vectors. Let $w_\mu$ be a positive
weight factor attached to each vector, indicating its importance. Let
$R(\abc)$ be the classical orthogonal rotation matrix \cite{goldstein}
for Euler angles $\abc$.  Then the cosine of the angle between Bob's
estimate of ${\bf e}^\mu_m$ and the true direction of that vector is

\beq \cos\omega_\mu=
 \sum_{m,n}R_{mn}(\abc)\,{\bf e}^\mu_m\,{\bf e}^\mu_n. \eeq
With the same notations as before, we have

\beq \6f(\abc)\9=\sum_\mu w_\mu\,\6\cos\omega_\mu\9=
 \sum_{m,n}\6R_{mn}(\abc)\9\,C_{mn}, \eeq
where

\beq C_{mn}=\sum_\mu w_\mu\,{\bf e}^\mu_m\,{\bf e}^\mu_n. \eeq
This is a positive matrix which depends only on the geometry of the
set of vectors whose transmission is requested. We can now diagonalize
$C_{mn}$ and write it in terms of three orthogonal vectors, possibly
with different weights. Therefore, no essentially new features follow
from considering more than three directions.

Finally, we note that all the above calculations, as well as those in
preceding works, assume that Alice and Bob have coordinate frames
with the same chirality (this can be checked locally by using weak
interactions). If the chiralities are opposite, then all the directions
inferred by Bob should be reversed (because directions are polar vectors
while spins are axial vectors).

In summary, we have shown that a single structureless quantum system
(a point mass in a Coulomb potential) can transfer information on the
orientation of a Cartesian coordinate system with arbitrary accuracy.
At first sight, this conclusion seems surprising. No spherically
symmetric classical object can achieve this result. Only those having
an asymmetric internal structure, such as an asymmetric rigid body,
can reliably transmit a Cartesian frame. However, a {\it classical\/}
Kepler orbit has two vectorial constants of the motion: the angular
momentum and the Lenz vector \cite{goldstein}. The Hamiltonian itself
is spherically symmetric, but each elliptic orbit (each solution of the
equations of motion) defines a unique Cartesian coordinate system. From
the correspondence principle which is valid for large quantum numbers,
we expect that there are sets of coefficients $a_{jn}$ such that the
wave function is concentrated in the vicinity of a classical elliptic
orbit, and thereby defines a Cartesian system. This problem is under
current investigation.

\section{Covariant quantum measurements may not be optimal}

In the preceding section, we discussed the transmission of a complete
Cartesian frame by a single quantum messenger, namely a hydrogen atom
(formally, a spinless particle in a Coulomb potential).  A similar
calculation was done by Bagan, Baig, and Mu\~noz-Tapia \cite{bbm}
(hereafter BBM), who were able to reach much higher values of $j$ by
sophisticated analytical methods.  There is an essential difference
between our work and that of BBM. We considered a single system,
while BBM took $N$ spins, and one irreducible representation for each
value $j$ of the total angular momentum. The maximum value is $j_{{\rm
max}}=N/2$, and then the mathematics are the same as for our Rydberg
state, with $j_{{\rm max}}= n-1$. However, as explained above, if
there are $N$ spins that can be sent independently, there is a better
method. Alice can use half of them to indicate one axis, and the other
half for another axis. This method is far more accurate, especially
if $N$ is large. From Eq.~(\ref{infi1}), the mean square error for
each axis is $5.783/(N/2)^2=23.13/N^2$, rather than $4/3N$ which is
the result with the method used by BBM \cite{bbm}.  Similar results
hold even for low values of $N$. Why is there such a discrepancy?

In all the works that were mentioned above, and in many other similar
ones, it was assumed that Holevo's method of covariant measurements
\cite{holevo} gave optimal results. That method considers the
case where Alice's signals are the orbit of a group $\cal G$, with
elements~$g$. Namely, if $|A\9$ is one of the signals, the others
are $|A_g\9=U(g)|A\9$, where $U(g)$ is a unitary representation of
the group element $g$.  The problem is to find optimal quantum
states for Alice's signals and Bob's detectors.  Originally, Holevo
considered only irreducible representations. Now we know that in some
cases reducible representations are preferable \cite{bbbmt,petra}.
However one then never needs to use more than one copy of each
irreducible representation in the reducible one.

We now turn our attention to Bob.  The mathematical representation
of his apparatus is as always a POVM, namely a resolution of identity
by a set of positive operators:

\beq  \sum_h E_h = \1,\eeq
where the label $h$ indicates the outcome of Bob's experiment. In the
case of {\it covariant\/} measurements, the labels $h$ run over all
the elements of the group $\cal G$ (with a suitable adjustment of the
notation in the case of continuous groups). Then the probability that
Bob's apparatus indicates group element $h$ when Alice sent a signal
$|A_g\9$ is

\beq P(h|g)=\6A_g|E_h|A_g\9. \label{Phg}\eeq
The method of covariant measurements further assumes that $E_h$ can
be written as

\beq E_h=|B_h\9\6B_h|,\label{cov} \eeq
where

\beq |B_h\9=U(h)|B\9. \eeq
Here, $|B\9$ is a fiducial vector for Bob (which has to be optimized)
and $U(h)$ is a representation (possibly a direct sum of irreducible
representations) of the same group $\cal G$ that Alice is using.

All this seems quite reasonable (and this indeed usually works well)
but, as the above example shows, this may not be the optimal method. In
the above example, Alice's signals $|A_g\9$, for all possible positions
of her axes, are $SO(3)$ rotations of a fiducial state $|A\9$ with
$j=0,...,n-1$. On the other hand Bob uses two separate POVMs, each
one testing only one half of the spins. Each one of these POVMs
also involves $SO(3)$, but with lower values of the maximal $j$.
Further discussion of this topic will appear in a forthcoming
publication \cite{jmo}.

\bigskip Work by AP was supported by the Gerard Swope Fund and the
Fund for Encouragement of Research. PFS was supported by a grant from
the Technion Graduate School.


\begin{thebibliography}{99}
\bibitem{maxwell}Owing to Gleason's theorem, any detection procedure
is mathematically represented by a POVM.  However, quantum mechanics is
only a theoretical framework and it does not tell how to actually build
the corresponding experimental setup. Likewise, Maxwell's equations do
not say how to build an electric motor, and thermodynamics does not tell
how to build steam engines.
\bibitem{polyhed}If $N\leq4$, an optimal POVM can have a finite number
of components, corresponding to the vertices of regular polyhedra. In
general, a POVM with a nonuniform distribution of directions on the unit
sphere can be converted into a uniform one by rotational symmetrization.
This does not affect the fidelity of transmission if the input is
uniform. We can therefore assume that the POVM distribution is uniform
without loss of generality. (See also ref.~16.)
\bibitem{cnvx}A convex combination of POVMs, such as $E_\mu=\sum
w_kE_{k\mu}$, with $w_k>0$ and $\sum w_k=1$, has the physical meaning
that each one of the sets $\{E_{k\mu}\}$ is chosen by the experimenter
with probability $w_k$.
\bibitem{mix}More generally we can take $|\theta,\phi\9=\sum_j
\sqrt{2j+1}\sum_m b_{jm}|j,m(\theta,\phi)\9$, with $\sum_m
|b_{jm}|^2=1$ for each $j$. Likewise, Alice's signal state (\ref{alice})
may include all $j$ and all $m$. In that case, it can be proved that
fidelity is optimal if a single value of $m$ is used.
\end{thebibliography}
\end{document}